\documentclass[10pt,preprint]{aastex}
\usepackage{graphicx}
\newcommand{\etal}{et al.~}
\def\gsim{\lower 2pt \hbox{$\, \buildrel {\scriptstyle >}\over
{\sim}\,$}}
\def\gtrsim{\lower 2pt \hbox{$\, \buildrel {\scriptstyle >}\over
{\sim}\,$}}
\def\lsim{\lower 2pt \hbox{$\, \buildrel {\scriptstyle <}\over
{\sim}\,$}}
\def\lesssim{\lower 2pt \hbox{$\, \buildrel {\scriptstyle <}\over
{ \sim}\,$}}
\def\rosat{{\sl ROSAT}}

\def\xmm{{\sl XMM-Newton}}

\def\suzaku{{\sl Suzaku}}
\def\chandra{{\sl Chandra}}

\def\ovii{O~{ VII}}
\def\ovi{O~{ VI}}

\pdfoutput=1
\begin{document}

\title{X-raying Galaxies: A Chandra Legacy}

\author{Q. Daniel Wang\altaffilmark{1}}
\altaffiltext{1}{Department of Astronomy, University of Massachusetts, Amherst, USA}


\begin{abstract}
This presentation reviews Chandra's major contribution to the understanding 
of nearby galaxies. After a brief summary on 
significant advances in characterizing various types of discrete X-ray 
sources, the presentation focuses on the global hot gas in and around
galaxies, especially normal ones like our own. 
The hot gas is a product of stellar and AGN feedback --- the 
least understood part in theories of galaxy formation and evolution. 
Chandra observations 
have led to the first characterization of the spatial, thermal, chemical, 
and kinetic properties of the gas in our Galaxy.  The 
gas is concentrated around the Galactic bulge and disk on scales of a few kpc. 
The column density of chemically-enriched hot gas on larger scales is 
at least an order magnitude smaller, indicating that it may not account for the
bulk of the missing baryon matter predicted for the Galactic halo
according to the standard cosmology. Similar results have also been
obtained for other nearby galaxies. The X-ray emission from hot gas is well 
correlated with the star formation rate and stellar mass, indicating that 
the heating is primarily due to the stellar feedback. However, 
the observed X-ray luminosity of the gas is typically 
less than a few percent of the feedback energy. Thus the bulk of the 
feedback (including injected heavy elements) is likely lost 
in galaxy-wide outflows. The results are compared with
simulations of the feedback to infer its dynamics and interplay with 
the circum-galactic medium, hence the evolution of galaxies.
\end{abstract}
  
\section{Introduction}

X-ray observations are playing an increasingly important role in 
the study of galaxies. With its arcsecond
spatial resolution, \chandra\ in particular has made a significant impact on 
our understanding of discrete X-ray source populations,
which mostly represent various stellar end-products [e.g., low- and high-mass
X-ray binaries (LMXBs and HMXBs) and supernova remnants (SNRs)] as well as 
active galactic nuclei (AGNs). The resolution also allows for a clean excision 
of such sources from the data so low-surface brightness emission
(e.g. from diffuse hot gas) can be mapped out.
An under-appreciated aspect of \chandra\ is its spectroscopic capability 
in the study of diffuse hot gas when the low- 
and high-energy grating instruments are used. Although the sensitivities
of the instruments are quite limited, the existing observations of 
a dozen or so bright objects (AGNs and LMXBs) have yielded 
data of high enough quality for unprecedented 
X-ray absorption line spectroscopic measurements of the 
global hot gas 
in and around our Galaxy. Useful constraints have also been obtained
on the overall content of hot gas around
galaxies within certain impact distances of the sight lines
toward the AGNs. These measurements, compared with physical models and
simulations of the hot gas, are shedding important insights on
its relationship to the feedback from stars and AGNs. These aspects of the 
\chandra's legacy are reviewed in the following.

\section{Discrete Sources}

The overall X-ray luminosity of a galaxy (except for a giant elliptical) 
is usually dominated by HMXBs and/or LMXBs. The luminosity functions 
of HMXBs and LMXBs are linearly scaled with 
the star formation rate and the stellar mass of a galaxy and are universal to an 
accuracy better than $\sim 50\%$ and $30\%$, respectively \cite{gri03,gil04}. The 
differential power law slope of the
function for HMXBs is $\approx 1.6$ over a broad range of 
log$(L_x) \sim 35.5-40.5$, where $L_x$ is the luminosity in
the 0.5-2 keV band. Particularly interesting are a large number of non-AGN 
(hyper)ultraluminous X-ray sources with log$(L_x)$ in the range of 
$\sim 39.5 - 41.5$ and observed typically in active star forming 
galaxies, suggesting the presence of either so-called intermediate-mass
black holes ($10 \lesssim M_{BH} \lesssim 10^5$) or sources apparently 
radiating well above the Eddington limit. 
The luminosity function shape for LMXBs is a bit more complicated,
having a slope of $\approx 1$ at log$(L_x) \lesssim 37.5$,
steepening gradually at higher luminosities and cutting off abruptly at 
log$(L_x) \sim 39.0-39.5$. Furthermore, the frequency of LMXBs per stellar mass
is substantially higher in globular clusters than in galaxy field 
(e.g., \cite{sar03}). This is attributed to the formation of LMXBs 
via close stellar encounters, which has also been proposed to 
account for an enhanced number of LMXBs in the 
dense inner bulge of M 31 \cite{vos07}. But it is not yet clear 
as to what fraction of field LMXBs was formed dynamically 
(e.g., \cite{whi02,vos09}). 

For the study of diffuse hot gas, it is important to minimize the confusion 
from point-like source contributions. A source detection
limit at least down to log$(L_x) \sim 37$ is highly desirable, 
which can be achieved with a {\sl Chandra} observation of 
a reasonably deep exposure for nearby galaxies 
($D\lesssim 20$ Mpc). The residual contribution 
from fainter HMXBs and LMXBs can then be estimated from their 
correlation with the star formation rate and stellar mass 
and subtracted from the data with little uncertainty. However, one still 
needs to be careful with Poisson fluctuations of sources just below 
the detection limit. Such fluctuations may significantly affect the 
reliability of X-ray morphological analysis of a galaxy.

In addition to the 
subtraction of relatively bright X-ray binaries, one also needs to
account for a significant (even dominant) stellar contribution 
from cataclysmic variables and coronally active stars, which are numerous, 
though individually faint. Very deep \chandra\ imaging of a region toward 
the Galactic bulge \cite{rev09} has resolved out more than 80\% of the 
background emission at energies of 6-7 keV, where the observed prominent 
Fe $6.7$-keV line was thought as the evidence for the presence of diffuse 
hot plasma at $T \sim 10^8$ K. This high-energy background emission is now 
shown to be entirely 
consistent with this collective stellar contribution in the
Galactic bulge/ridge. It should be noted, however, that the resolved fraction
is much smaller at lower energies ($\sim 50\%$ at $\lesssim 4$ keV), which 
may be considered as an indication for the presence of diffuse hot gas at 
much lower temperatures. 
The stellar contribution is
unresolved for external galaxies, even nearby ones. Fortunately, the average 
X-ray spectrum and specific luminosity (per stellar mass) of the contribution 
have been calibrated, based on the 
\chandra\ observations of M32, which is too light to hold significant
amount of diffuse hot gas), together with the direct detection of
stellar X-ray sources in the solar neighborhood \cite{rev08}. The contribution
can be readily included in a spectral analysis of the ``diffuse''
X-ray emission of a galaxy. In an imaging analysis, one may subtract
the contribution scaled according to the stellar 
mass distribution (e.g., traced by the near-IR K-band intensity of a galaxy).

\begin{figure}[!tbh]
\centerline{\includegraphics[width=0.5\textwidth]{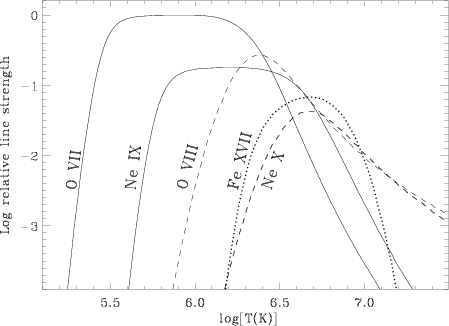}}
\caption{ Relative line strength (ionization fraction 
times oscillation strength) of the K$\alpha$ lines of key species that
trace gas in the temperature range of $10^{5.5}-10^{6.5}$ K \cite{gna07}. 
} 
\label{fig:f1}
\end{figure}

\section{Global hot gas in and around our Galaxy}

The properties of the global hot gas on scales comparable to the size of the
Galaxy remained largely unknown until recently. 
Before \chandra, we did have various broad-band 
X-ray emission surveys of the sky such as the 
one made by \rosat\ in the 0.1-2.4 keV range, which is sensitive to the 
hot gas \cite{sno97}. But such a survey
alone cannot be used to directly determine the origin of the X-ray emission,
which carries little distance information. 
The interpretation of the emission depends sensitively on the assumed 
cool (X-ray-absorbing) and hot gas distributions. Spectroscopic information 
on the X-ray emission has been obtained from rocket experiments \cite{mcc02}
and more recently from \suzaku\ observations, but only for a number of 
sample regions (e.g., \cite{yos09}). There are
also large uncertainties in the contributions from faint discrete sources and 
other irrelevant phenomena such as solar wind charge exchange (SWCX) 
to the emission. 

A breakthrough for the study of the global hot gas has been made
from the use of the X-ray absorption line spectroscopy 
(e.g., \cite{fut04, wan05,wil05,yw05,yw06,yw07a,yw07b,yao08,yao09}). 
While absorption line spectroscopy is commonly used in optical and UV studies
of the interstellar medium (ISM), this technique became feasible in 
the soft X-ray regime only with grating spectra from \chandra\ and \xmm.
Unlike the emission, which is sensitive 
to the density structure, absorption
lines produced by ions such as O VII, O VIII, and Ne IX (Fig.~1~\ref{fig:f1})
directly probe their column densities, 
which are proportional to the mass of the hot gas. 
The relative strengths of such absorption lines give direct diagnostics
of the thermal, chemical and/or kinetic properties of the hot gas. 
Although the absorption lines are rarely resolved in the spectra (with
a resolution of $\sim 400-500 {\rm~km~s^{-1}}$ FWHM), the velocity
dispersion of the hot gas can be derived from the relative line
saturation of different transitions of same ion species (e.g.,
\ovii\ K$\alpha$ vs. K$\beta$). Because the K-shell transitions of  
carbon through iron
are all in the X-ray regime, the same technique can be used to study the ISM in
essentially all phases (cold, warm, and hot) and forms (atomic, molecular,
and dust grain; \cite{yw06}). Furthermore,
the measurements are insensitive to the photo-electric absorption 
by the cool ISM  ($kT \lesssim 10^4$ K) and to the SWCX.
Therefore, the X-ray absorption line spectroscopy 
allows us to probe the global ISM unbiasedly along a sight line.
The effectiveness of the technique can be further enhanced
when multiple sight lines are analyzed jointly (e.g., \cite{yw07b}) 
and/or emission data are included (e.g., \cite{yao08}). We can then infer 
differential properties of hot gas between sight lines of
different depths or directions and/or estimate the pathlength and density of the hot gas. The application of this X-ray tomography, though only to a very 
limited number of sight lines so far, has led
to the first characterization of the global hot gas:

\begin{itemize}
\item The spatial distribution of the gas at the solar neighborhood 
can be characterized by a thick Galactic 
disk with a density scale height of $\sim 2$ kpc
 (e.g., \cite{yw05,yw07a,yao09}), comparable to those measured for 
\ovi\ (from far-UV absorption line observations) and free electrons 
(from pulsar dispersion measures). The density is enhanced
toward the inner region of the Galaxy, indicating the presence of
a Galactic bulge component of the hot gas \cite{yw05,yw06,yw07b}.
A temperature increase is also observed from the disk 
($\sim 10^{6.2}$ K) to the bulge ($\sim 10^{6.4}$ K; e.g., 
\cite{yw05,yw06,yw07a,yw07b,yao09}).

\item The velocity dispersion increases from $\sim 60 
{\rm~km~s^{-1}}$ at the solar neighborhood to $\sim 300 {\rm~km~s^{-1}}$ 
toward the inner region of the Galaxy (with 90\% uncertainties up to
$\sim 40\%$; e.g., \cite{yw06,yw07a}), indicating a significant nonthermal
(e.g., turbulent) velocity contribution to the line broadening
(cf. the thermal broadening $\sim 40 {\rm~km~s^{-1}}$).

\item The metal abundances (O/Fe, Fe/Ne, and O/Ne) are about solar, although
there are lines of evidence for the depletion of O and Fe (by a factor
of $\sim 2$) at the solar neighborhood (e.g., \cite{yw06,yao09}), assuming the
abundances in \cite{ag89}.
.

\item No evidence is yet found for a large-scale 
X-ray-absorbing hot circum-galactic medium (CGM) with a 95\% 
upper limit to the \ovii\ column density 
$N_{OVII} \sim 3 \times10^{15} {\rm~cm^{-2}}$ for regions $\gtrsim 10$ kpc away 
from the Galactic disk/bulge \cite{yao08} and $\sim 1 \times10^{15} 
{\rm~cm^{-2}}$ for $\gtrsim 50$ kpc \cite{yao09}. This low column density
of the hot CGM or the local hot intragroup medium is consistent with
the lack of evidence for enhanced X-ray line absorptions along sight lines
near the major axis of the Local Group \cite{bre07}.
\end{itemize}

This basic characterization demonstrates the potential of using the X-ray 
tomography to greatly advance our understanding of 
the global hot gas in and around the Galaxy.

\section{Global hot gas in and around other galaxies}

\begin{figure}[bht]
\centerline{\includegraphics[width=0.9\textwidth]{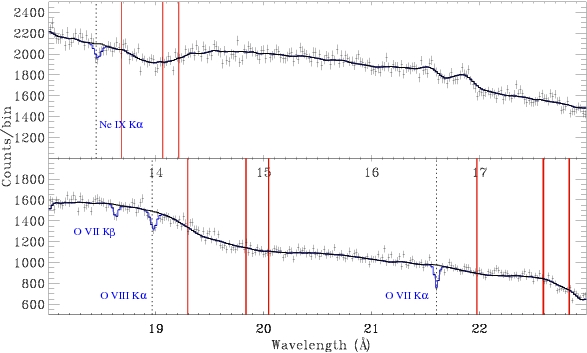}}
\caption{ Part of the \chandra\ grating spectrum of PKS 2155-304
 with a total exposure of 1 Ms. Various absorption lines (marked in 
blue) are detected and are produced by hot gas in and
around the Galaxy (zero redshift). The red-shaded regions enclose 
spectral ranges of the same lines for three red-shifted groups of
galaxies located within the 500 kpc impact distance of the AGN sight line.
} 
\label{fig:f2}
\end{figure}

The X-ray absorption line spectroscopy can also be used to probe 
the hot CGM around other galaxies. Yao et al. have recently 
studied the hot CGM along the sight lines toward luminous AGNs \cite{yao10}. 
As an example, Fig.~2
shows a \chandra\ grating spectrum of PKS 2155-304.
No absorption line is apparent at the 
redshifts of the three groups of galaxies with the
impact distances smaller than 500 kpc of the
sight line (in particular, the group at the highest redshift contains two galaxies
brighter than $L_*$). The same
is true when the spectrum is folded according to the redshifts
of the individual groups. To increase the counting statistics further, 
Yao et al. have stacked spectra from the \chandra\ grating observations 
of eight AGNs with such intervening galaxies or groups.
No significant absorption is detected for any of these 
individual systems or in the final stacked spectrum
with a total equivalent exposure of about 10 Ms! 
Upper limits to the mean column densities of various ion species per galaxy
or group
are then estimated. In particular, they find that 
$N_{\rm OVII}\le 6 \times 10^{14}~{\rm cm^{-2}}$, 
consistent with the constraints for the CGM around our Galaxy.
They have estimated the
total mass contained in the CGM as 
$M_{\rm CGM}\lsim 0.6 \times(\frac{0.5}{f_{\rm
      OVII}})\times(\frac{0.3A_\odot}{A})\times(\frac{R}{500~
  {\rm kpc}})^2\times10^{11}M_\odot$,  
where $f_{\rm OVII}$, $A$, and $R$ are the ionization fraction of 
\ovii, metal abundance, and the radius of the hot CGM, respectively.
This is in contrast to the expected baryon mass
$\gsim2\times10^{11}~M_\odot$ for the halo of a Milky Way-type galaxy or
a typical galaxy group \cite{mcg09}.
Thus the bulk of the CGM unlikely resides 
in such a chemically enriched warm-hot phase 
at temperatures ranging from $10^{5.5}-10^{6.5}$ K (Fig.~1~\ref{fig:f1}),
which our X-ray absorption line spectroscopy is sensitive to. 
This conclusion has strong implications for understanding the 
accumulated effect of the stellar and AGN feedback on the galactic ecosystem 
(see the discussion section). 

To study the effect of ongoing stellar and AGN feedback, 
one can map out diffuse 
X-ray emission from hot gas in and around nearby galaxies of various
masses and star formation rates. 
Much attention has been paid to the feedback in starburst and 
massive elliptical galaxies, which are relatively bright in
diffuse X-ray emission. \chandra\ observations have shown convincingly
that the AGN feedback is important in shaping the morphology and thermal 
evolution of hot gas in massive elliptical galaxies,
particularly those at centers of galaxy groups and clusters 
(\cite{mn07} and references therein). The asymmetry in the global
diffuse X-ray morphology is correlated with radio and X-ray 
luminosities of AGNs in elliptical galaxies, even in rather X-ray-faint
ones \cite{die08}. This calls into question the hydrostatic assumption commonly used
in order to infer the gravitational mass distribution in such galaxies.
Nevertheless, the hydrostatic assumption may hold approximately 
for hot gas around the central supermassive black holes (SMBHs), if they are in a sufficiently quiescent state. 
The SMBH masses may then be measured from 
spatially resolved X-ray spectroscopy of the hot gas. 
Humphrey et al. have made such mass measurements for four SMBHs
with \chandra\ data \cite{hum09}. Three of them 
already have mass determinations from the kinematics of either 
stars or a central gas disk. It is encouraging to find a good agreement 
between the measurements using the different methods. From this
agreement, they further infer that no more than $\sim 10\%-20\%$ of the 
ISM pressure around the SMBHs should be nonthermal.

The feedback in nuclear starburst galaxies is manifested in the so-called
galactic superwinds driven by the mechanical energy 
injection from fast stellar winds and supernovae (SNe) of massive stars
 (e.g., \cite{str04a,str04b,sh09}). The observed soft X-ray emission 
from a superwind typically has an elongated morphology along the minor axis 
of such a galaxy and 
is correlated well with extraplanar H$\alpha$-emitting features. This indicates
that the detected hot gas arises primarily from the interaction
between the superwind and cool gas. The superwind itself, believed to 
be very hot and low in density, is much difficult to detect.
From a detailed comparison between \chandra\ data and hydrodynamic 
simulations, Strickland \& Heckman infer that the superwind 
of M82 has a mean temperature of $3-8 \times 10^7$ K and a mass
outflowing rate of $\sim 2 {\rm~M_\odot~yr^{-1}}$ \cite{sh09}. 
Such energetic superwinds with little radiative energy loss must
have profound effects on the large-scale CGM (e.g., \cite{str04b}).

Recent X-ray observations have further shown the importance of the feedback in 
understanding even ``normal'' intermediate-mass galaxies 
(similar to the Milky Way and M31; e.g.,
\cite{str04a,str04b,wan03,tyl03,doa04,tul06a,tul06b,lw07,lij08,bg08,yam09}). 
\chandra, in particular, has unambiguously detected diffuse hot gas
in and around normal disk galaxies. The total X-ray luminosity of the gas
is well correlated with the star formation rate for such galaxies. The diffuse soft X-ray emission is 
shown to be strongly enhanced in recent star forming regions or spiral 
arms within an individual galaxy viewed face-on 
and is only slightly more diffuse than H$\alpha$ emission
(e.g., \cite{tyl03,doa04}). This narrow appearance of spiral arms in X-ray 
conflicts the expectation from population synthesis models:
the mechanical energy output rate from SNe should be nearly constant over
a time period that is up to 10 times longer than the lifetime of 
massive ionizing stars. This means that SNe in inter-arm regions, where ISM density
is generally low, must produce less soft X-ray emission than those in the arms. 
When disk galaxies are observed in an inclined angle (e.g., Fig.~3),
the soft X-ray emission tends to appear as plumes,
most likely representing blown-out hot gas heated in   
recent massive star forming regions and galactic spheroids. The cleanest 
perspectives of the extraplanar hot gas are obtained from the observations
of edge-on disk galaxies such as NGC 891 \cite{str04a} and NGC 5775 
\cite{lij08}. The observed diffuse X-ray emission typically does
not extend significantly more than a few kpc away from the galactic 
disks. A claimed detection of the 
emission around the edge-on spiral NGC 5746 on larger scales was later proved
to be due to an instrumental artifact \cite{ras09}. Complementary
observations from {\sl XMM-Newton} and {\sl Suzaku} give 
consistent results and provide improved spectral information on the hot gas 
(e.g., 
\cite{tul06a,tul06b,yam09,liz06,liz07}). The morphology of the X-ray emission as well as 
its correlation with the star formation rate clearly shows that the extraplanar
hot gas is primarily heated by the stellar feedback. 
In fact, the cooling of the hot gas accounts for only a small fraction
of the expected energy input from massive stars alone (typically no more
than a few \%). 

\begin{figure}[tbh]
\centerline{\includegraphics[width=0.8\textwidth]{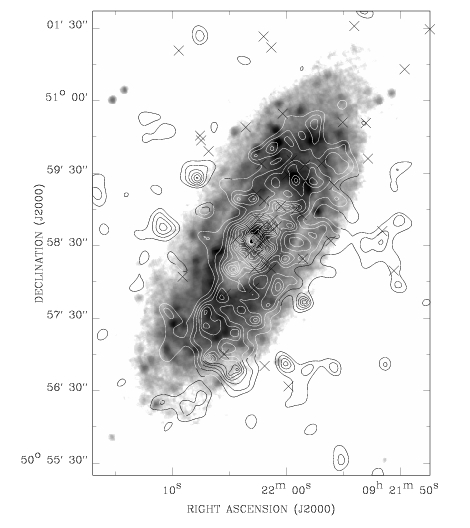}}
\caption{ \chandra\ diffuse 0.5--1.5~keV intensity contours, 
overlaid on a {\sl Spitzer} mid-IR image of the Sb galaxy NGC~2841. The 
one-sided morphology of
the diffuse X-ray emission apparently represents outflows of hot gas 
from the tilted galactic disk with its northeast edge closer to us. 
The emission from the back side is largely absorbed by the cool gas in the disk.
The {\sl crosses} mark the positions of excised discrete X-ray sources.
} 
\label{fig:f3}
\end{figure}

The best example of mapping out hot gas in a galactic stellar
bulge (or spheroid) is the discovery of an apparent hot gas outflow
from the M31 bulge \cite{lw07,bg08}. This outflow
with a 0.5-2 keV luminosity of $\sim 2 \times 10^{38} {\rm~erg~s^{-1}}$
is driven apparently by the feedback from evolved stars in form of 
stellar mass loss and (Type) Ia SNe, because there is no evidence for an AGN
or recent massive star formation in the bulge.
The bipolar morphology of the truly diffuse soft X-ray emission 
further indicates that the outflow is probably  
influenced by the presence of strong vertical magnetic field, same as 
that observed in the central region of
our Galaxy. The spectrum of the emission can be characterized by 
an optically-thin thermal plasma with a temperature 
of $\sim 3 \times 10^6$ K \cite{lw07,bg08,liz09}. 
Similar analyses have been carried out so far only for a couple of
other early-type galaxies (NGC 3379, \cite{tri08}; NGC 5866,  \cite{lij09}). 
The faint stellar contribution, which has a significantly different
spectral shape from LMXBs', still needs to be carefully accounted for in 
the analysis of other galaxies. In any case, it is clear that the
diffuse X-ray luminosity accounts for at most a few \% of the energy input 
from Ia SNe alone in a normal early-type galaxy.
Where does the missing feedback go?

Another remarkable puzzle about the diffuse hot gas in elliptical galaxies is the 
apparent low metal abundances. As inferred from X-ray spectral fits,
the abundances are typically sub-solar for low- and intermediate-mass galaxies
with log($L_x) \lesssim 41$ to about solar for more massive ones 
(e.g., \cite{hb06, jij09}), substantially less than what
are expected from the Ia SN enrichment (see the discussion section).
Furthermore, a number of galaxies show a significant
iron abundance drop toward their central regions, where the stellar 
feedback should be the strongest. Examples of this abundance drop 
include M87 \cite{gm02}, 
NGC 4472, NGC 5846 \cite{buo00}, and NGC 5044 \cite{buo03}. The drop and the low abundance, if intrinsic, 
would then indicate that only a small portion of metals produced 
by Ia SNe is observed; the rest is either expelled or in a 
state that the present X-ray data are not sensitive to. 

In summary, the following are the key discoveries made with \chandra\ 
observations of diffuse hot gas in and around normal galaxies:
\begin{itemize}
\item The hot gas generally consists of two components: 1) the disk component, which is sensitive to the star formation rate 
per unit area (e.g., \cite{str04b, lij08, dav06, liz07}); 
2) the bulge component, the amount of which is proportional to the stellar mass.
Significant diffuse soft X-ray emission is detected only within 
$\sim 10$ kpc of the disk and bulge of a galaxy. 

\item The gas typically has a mean characteristic temperature of 
$\sim 10^{6.3} K$, although there is evidence for a considerably high
temperature component (up to $\gtrsim 10^7$ K) in and around galactic 
disks with active star formation \cite{doa04,lij08}. This high-temperature
component may represent a significant galactic hot gaseous outflow. The 
interaction of this outflow with cool gas (via shocks, dynamic mixing, and/or
charge exchanges) is responsible for much of 
the low-temperature one (e.g., \cite{sh09,lij08}).

\item The cooling rate, as inferred from the diffuse X-ray emission, 
is insignificant, compared to the expected stellar feedback alone. The bulk of
the feedback is missing, especially in inter-arm regions and in galactic
spheroids.

\item The hot CGM seems to have a very low density
and may not account for the missing baryon matter expected in individual
galactic dark matter halos. 

\end{itemize}

While the above results are
generally consistent with those found for our own Galaxy, 
the studies of nearby galaxies also show intriguing sub-structures in 
the diffuse hot gas distributions (e.g., the apparent 
positive radial temperature and metal abundance gradients in elliptical 
galaxies).

\section{Discussion}

As shown above, the diffuse soft X-ray emission 
observed in and around galaxies clearly traces the global hot gas that is heated by the stellar and possibly AGN feedback. By comparing the X-ray 
observations directly with physical models
and/or simulations of the feedback, we may further study its dynamics, which
is so far hardly known. The following discussion is concentrated on 
the modeling of the feedback in 
galactic spheroids for their relative simplicity. The presence of substantial
cool gas and star formation, as in typical disk galaxies, would 
certainly add complications. 

\subsection{Comparison of feedback models with X-ray observations}

The missing feedback problem is 
particularly acute in so-called low $L_X/L_K$ spheroid-dominated galaxies.
Empirically, the Ia SN rate is $\approx 0.019 {\rm~SN~yr^{-1}}) 
10^{- 0.42(B-K)} [L_K/(10^{10}L_{\odot K})]$,
where $ L_{K}$ is the K-band luminosity of a galaxy \cite{man06}. 
Adopting the color index $B-K \approx 4$ for a 
typical spheroid, we estimate the total mechanical energy
injection from Ia SNe (assuming $10^{51} {\rm~ergs}$ each) is 
$L_{Ia} \approx (1.3 \times 10^{40} {\rm~ergs~s^{-1}}) 
[L_K/(10^{10}L_{\odot K})]$. The mechanical energy 
release of an AGN can also be estimated empirically \cite{bes06, dav06} as
$L_{AGN} = (1.1 \times 10^{39} {\rm~erg~s^{-1}}) 
[L_K/(10^{10} L_{K,\odot})]^2.$
Therefore, averaged over the time, Ia SNe are energetically more important than
the AGN in a galactic spheroid with $L_K \lesssim 10^{11}L_{K,\odot}$.
Furthermore, SN blastwaves provide a natural distributed heating mechanism for 
hot gas in galactic spheroids \cite{tw05}. Additional continuous heating is 
expected from converting the kinetic energy of stellar mass loss from
randomly moving evolved stars to the thermal energy.
On the other hand, the AGN feedback, likely occurring in 
bursts with certain preferential directions (e.g., in form of jets), 
can occasionally result in significant 
disturbances in global hot gas distributions, as reflected 
by the asymmetric X-ray morphologies observed in some elliptical galaxies 
(e.g., \cite{die08}). 
It is not yet clear as to what fraction of the AGN feedback energy is converted
into the heating of the hot gas observed. 
In general, the lack of distributed cool gas in galactic spheroids
makes it difficult to convert and release the thermal energy into radiation
in wavelength bands other than the X-ray. 

Not only the mechanical 
energy, the gas mass from the stars and Ia SNe, especially heavy
elements, is missing as well. The mass injection rate 
is expected to be $0.026[L_K/(10^{10} L_{K,\odot})] 
M_\odot {\rm~yr^{-1}}$ with a
mean iron abundance $Z_{Fe} \approx Z_{*,Fe}+ 4(M_{Fe}$$/0.7$$ M_\odot)$,
where $Z_{*,Fe}$ is the iron abundance of the stars while
$M_{Fe}$ is the iron mass yield per Ia SN (e.g., 
\cite{man06,kna92,cio91}). The above empirical estimates of the
energy and mass feedback rates, which should be accurate within a 
factor of $\sim 2$, are typically a factor of $\sim 10^2$ greater than what
are inferred from the diffuse X-ray emission. 
Naturally, one would expect that the missing feedback
is gone with a wind (outflow), spheroid-wide or even galaxy-wide.

The notion that Ia SNe may drive galactic winds is not new 
(e.g., \cite{cio91, mb71, bre80}). But could such 
winds explain the diffuse X-ray emission of galactic spheroids?  It is easy to construct a 1-D 
steady-state supersonic wind model, assuming that the specific energy
of the feedback (per mass) of a galaxy is large enough 
to overcome its gravitational bounding and that the CGM pressure
(thermal or ram) is negligible (e.g., \cite{tan09b}). This supersonic wind model depends 
primarily on two feedback parameters: the integrated energy 
and mass input rates. 
However, the model in general fails miserably: it predicts a 
too low luminosity (by a factor of $\sim 10^2$), a too high 
temperature (a factor of a few), and a too steep 
radial intensity profile to be consistent with \chandra\ observations 
of low $L_X/L_B$ galactic spheroids, in particular
those in M31 and M104 \cite{lw07,liz07}. Only 
few very low-mass and gas-poor spheroids, hence very faint in X-ray emission,
still seem to be consistent with the 1-D wind model (e.g., 
\cite{tri08}). 

\begin{figure}[tbh]
\centerline{\includegraphics[width=0.9\textwidth,angle=0]{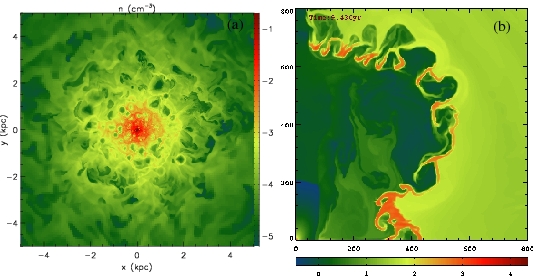}}
\caption{ Sample snapshots of hydrodynamic simulations of the stellar feedback.
{\bf (a)} the 3-D
simulated gas density distribution in an M31-like galactic spheroid. The slice 
is cut near the spheroid center and the units are in atoms~cm$^{-3}$, 
logarithmically.
{\bf (b)} Simulated 2-D large-scale gas density distribution
in the $r-z$ plane (in units of kpc).
A nearly vertical magnetic field, similar to what is observed in
the inter-cloud medium of the Galactic center, is included in 
the simulation to 
test its confinement effect on the spheroid wind; a reverse-shock is
clearly visible. Also apparent are instabilities at the contact 
discontinuity between the shocked spheroid wind gas and
the ejected materials from the initial starburst. The density is scaled
in units of M$_{\odot} {\rm~kpc^{-3}}$.
} 
\label{fig:f4}
\end{figure}

To compare with the X-ray emission, in fact one needs to account for 
3-D effects. X-ray emission is proportional to the 
emission measure and is thus sensitive to the detailed structure of hot gas
in a galactic spheroid. To realistically generate the inhomogeneity in the
heating and chemical enrichment processes, Tang \& Wang have developed a 
scheme to embed adaptively selected 1-D SNR seeds in 
3-D spheroid-wide simulations of supersonic winds or subsonic outflows
(e.g., \cite{tw05,tw09}). These 3-D simulations, reaching a resolution down to adaptively 
refined scales of a few pc (e.g., Fig.~\ref{fig:f4}a), show several important 3-D effects \cite{tan09b}:
\begin{itemize}
\item Soft X-ray emission arises primarily from relatively low temperature and low abundance gas shells associated with SN blastwaves. The inhomogeneity in the gas density and temperature substantially alters the spectral shape and 
leads to artificially lower metal abundances (by a factor of a few) 
in a spectral fit with a simplistic thermal plasma model. 
\item Reverse shock-heated SN ejecta, driven by its 
large buoyancy, quickly reaches a substantially higher outward velocity 
than the ambient medium. The ejecta is gradually and dynamically mixed with 
the medium at large galactic radii and is also slowly diluted and cooled by 
{\sl insitu} mass injection from evolved stars. These processes together 
naturally result in the observed positive gradient in the average radial 
iron abundance distribution of the hot gas, even if mass-weighted.
\item The average 1-D and 3-D simulations give substantially different radial 
temperature profiles; the 
inner temperature gradient in the 3-D simulation is positive, mimicking a 
``cooling flow''. (Thus the study of nearby galactic spheroids 
may provide important insights into the behavior of the intragroup
or intracluster gas around central galaxies, for which sporadic AGN energy
injections somewhat mimic the SN heating considered here.). 
\item The inhomogeneity also enhances the diffuse X-ray luminosity by a 
factor of a few in the supersonic wind case and more in a subsonic outflow.
In addition, a subsonic outflow can have a rather flat radial X-ray 
intensity distribution. 
\item  The dynamics and hence the luminosity of a subsonic flow further
depend on the spheroid star formation and feedback history. 
This dependence, or the resultant luminosity variance, may then explain 
the observed large dispersion in the X-ray to K-band luminosity ratios of 
elliptical galaxies. 
\end{itemize}

All considered, subsonic outflows appear to be most consistent 
with the X-ray observations of diffuse hot gas in typical intermediate-mass
galactic spheroids and elliptical galaxies. 

\subsection{The interplay between the feedback and galaxy evolution}

Whether a galactic outflow is supersonic or subsonic depends not only on
the specific energy of the ongoing feedback, but also on the properties of the
CGM, which  is a result of the past interplay between the feedback and
the accretion of a galaxy or a group of galaxies from the intergalactic
medium (IGM). Tang et al. have illustrated how this 
interplay may work, based on several 1-D hydrodynamic simulations in the 
context of galaxy formation and evolution \cite{tan09a}. 
They approximate the feedback history as having two distinct  
phases: (1) an early starburst during the spheroid formation (e.g., as a result
of rapid galaxy mergers) and (2) a subsequent long-lasting and slowly declining 
injection of mass and energy from evolved low-mass stars.
An energetic outward blastwave is initiated by the starburst (including
the quasar/AGN phase) and is sustained by the long-lasting stellar feedback.
Even for a small galactic spheroid such as the one in the Milky Way, 
this blastwave may heat
up the CGM on scales beyond the present virial radius,
thus the gas accretion from the IGM into the galactic halo could 
be largely reduced (see also \cite{kim09} 
for similar results from 3-D cosmological structure formation simulations). 
The long-lasting stellar feedback initially drives a galactic spheroid wind
(Fig.~\ref{fig:f4}b). As the mass and energy injection
decreases with time, the feedback may evolve into a subsonic and quasi-stable 
outflow. This feedback/CGM interplay scenario provides a natural explanation to 
various observed phenomena: 

\begin{itemize}
\item It solves the missing feedback problem 
as discussed earlier; the energy is consumed in maintaining the hot 
CGM and preventing it from fast cooling. 
\item The very low density of the CGM explains
the dearth of the chemically enriched hot gas observed around galaxies; 
much of the CGM or the intragroup gas has been pushed away to
larger scales (see also \cite{opp08}). 
\item The predicted 
high temperature is 
consistent with observations of the large-scale hot intragroup medium, 
which seems always higher than $10^{6.5}$ K \cite{sun09}. 
\item The cooling of the material ejected early (e.g., during 
the initial starbursts; \cite{tan09a}) provides a natural mechanism for 
high-velocity clouds with moderate metal abundances 
(Fig.~\ref{fig:f4}b). Such
clouds may be seen in 21 cm line emission and in various absorption
lines (e.g., Ly$\alpha$, Mg II, Si III; \cite{sto06,ws09,shu09}). 
O VI line absorptions can then arise from either photo-ionization
of the clouds or collisional ionization at their interfaces with the pervasive
hot CGM. 
\end{itemize}

\section{Summary}

Based on complementary high-resolution imaging and spectroscopic observations from 
\chandra, we have learned a great deal about various 
high-energy phenomena and processes in galaxies:

\begin{itemize}
\item The luminosity functions and their dependence
on the stellar mass and star formation rate are nearly universal for
essentially all major types of discrete X-ray sources in galaxies. 
\item A very hot component of the global ISM in the Galaxy has been revealed, 
and its global spatial, thermal, chemical, and kinetic properties 
have been characterized.
\item Stringent upper limits have been obtained to the contents 
of the chemically-enriched CGM around our Galaxy and around other galaxies/groups
along the sight lines toward several luminous AGNs.
\item Detailed structures of the superwinds
emanated from starburst galaxies are resolved, tracing strong interplay between
the very hot outflowing gas and the presence of cool gas.
\item Sporadic energy injections from AGNs are shown to play an important role in shaping the global hot gas in and around elliptical galaxies. 
\item Truly diffuse hot gas has also 
been mapped out for a few nearby normal galaxies, indicating that ongoing 
stellar feedback may play an important role in regulating the galactic 
eco-systems.
\end{itemize}

In conclusion, the existing work has demonstrated the power of 
\chandra\ observations in probing the stellar and AGN feedback and its effect 
on galaxy evolution as well as inventorying various kinds of 
high-energy sources in galaxies.





\begin{acknowledgments}
I thank the referee for constructive comments and the organizers 
of the Chandra's First Decade of Discovery 
conference for the invitation to give the talk that this paper is based on 
and am grateful to my students and collaborators for their contributions 
to the work described above, particularly 
Yangsen Yao who helped to produce Figures 1 and 2. The work is 
partly supported by NASA/CXC under grants G08-9088B and G08-9047A.
\end{acknowledgments}





\end{document}